\documentclass[twocolumn,amsmath,amssymb]{revtex4}
\usepackage[dvips]{graphicx}
\begin{document}

\def\lsim{\mathrel{\rlap{\lower4pt\hbox{\hskip1pt$\sim$}}
    \raise1pt\hbox{$<$}}}
\def\gsim{\mathrel{\rlap{\lower4pt\hbox{\hskip1pt$\sim$}}
    \raise1pt\hbox{$>$}}} 
\newcommand{\vev}[1]{ \left\langle {#1} \right\rangle }
\newcommand{\bra}[1]{ \langle {#1} | }
\newcommand{\ket}[1]{ | {#1} \rangle }
\newcommand{\ev}{ {\rm eV} }
\newcommand{\kev}{{\rm keV}}
\newcommand{\mev}{{\rm MeV}}
\newcommand{\gev}{{\rm GeV}}
\newcommand{\tev}{{\rm TeV}}
\newcommand{\mpl}{$M_{Pl}$}
\newcommand{\mw}{$M_{W}$}
\newcommand{\Ft}{F_{T}}
\newcommand{\Zparity}{\mathbb{Z}_2}
\newcommand{\BLambda}{\boldsymbol{\lambda}}
\newcommand{\be}{\begin{eqnarray}}
\newcommand{\ee}{\end{eqnarray}}

\title{Proposal for Higgs and Superpartner Searches at the LHCb Experiment}
\author{David E. Kaplan and Keith Rehermann}
\date{\today}

\begin{abstract}
  The spectrum of supersymmetric theories with R-parity violation are much more weakly
  constrained than that of supersymmetric theories with a stable neutralino.  We investigate the
  signatures of supersymmetry at the LHCb experiment in the region of
  parameter space where the neutralino decay leaves a displaced
  vertex.  We find sensitivity to squark production up to squark
  masses of order 1 TeV.  We note that if the Higgs decays to
  neutralinos in this scenario, LHCb  should see the lightest Higgs boson before
  ATLAS and CMS.
\end{abstract}

\maketitle

\section{Introduction}
Supersymmetry, and more specifically the minimal supersymmetric
standard  model (MSSM) \cite{Dimopoulos:1981zb} is a possible solution
to the gauge-hierarchy problem and a favorite model governing
physics  above 100 GeV.  Quantum corrections to the electroweak
breaking scale are proportional to the superpartner masses, and thus
one expects the MSSM spectrum to lie around the $Z$ mass.  Any
significant  deviation thereof necessitates a fine-tuning of parameters.

Minimal Supergravity (mSUGRA) is the most studied realization of the MSSM.  It assigns 
universal masses to all scalars and to all
gauginos at the scale $10^{16}$ GeV.  Experimental bounds on the mSUGRA spectrum demand that 
the model is tuned to the per cent level
\cite{Giudice:2006sn}.  Squarks and gluinos have lower bounds
around 300-400 GeV \cite{D0qg}, well above the $Z$ mass.  The corrections necessary to generate a
Higgs mass consistent with direct LEP bounds \cite{Schael:2006cr}
requires even more fine-tuning. 
The tight constraints on mSUGRA compel us to study more natural --
\textit{i.e}.  less fine-tuned -- models of supersymmetry.

The most general superpotential with the MSSM field content includes 
lepton and baryon number violating terms \cite{Weinberg:1981wj}
\begin{equation}
\lambda_{ijk}L_{i}L_{j}E^{c}_{k} + \lambda_{ijk}^{'}L_{i}Q_{j}D^{c}_{k} +
\lambda_{ijk}^{''}U^{c}_{i}D^{c}_{j}D^{c}_{k} ,
\label{eq:BNV}
\end{equation}
where $ijk$ are flavor indices.  
Bounds on proton decay severely constrain the combination of baryon
and lepton number violation.  Separately, however, they are much
more weakly constrained.  Throughout the following we restrict our
attention to the baryon number violating operators, the $\lambda^{''}$
terms.  This choice is motivated by the interesting and challenging
phenomenology it produces.  The bounds for squark masses in this scenario
are typically below 100 GeV \cite{Achard:2001ek}; some particles, such
as the gluino and lightest sbottom, do not have published bounds above
$\sim 10$ GeV in regions of parameter space
\cite{Janot:2004cy}.  Moreover, the bound on the Higgs can be below the
Z mass when decays to neutralinos are kinematically allowed \cite{Carpenter:2006hs}.

The phenomenologically interesting feature of this model is that the
lightest superpartner - taken to be a neutralino - is unstable.  It
decays to three quarks.  With regard to supersymmetry at the LHC,
the signals are changed significantly -- missing energy signals are
largely absent, and the number of isolated leptons is reduced due to
increased soft jet production \cite{Baer:1996wa}.  The ATLAS and CMS 
experiments \cite{AandC} will
have weakened sensitivity to this scenario because their triggers are
designed to exploit missing transverse energy and isolated leptons.  In the case
of squark or gluino production the associated jets should pass the triggers at
ATLAS and CMS, however hard jets are typically pre-scaled by a large factor that would significantly
reduce the effective luminosity \cite{Prescale}.  Even if the trigger issue is solved, it is not
clear that pure multi-jet events coming from this new physics can be
seen above the (unknown) QCD background.  Yet
more worrisome is the Higgs decay.  For decays to a final state of six
soft jets, no obvious search strategy presents itself, while the
standard searches are made more difficult with the reduced branching ratios.

The lightest neutralino has a macroscopic decay length in broad
regions of parameter space.  While neither ATLAS nor CMS currently
employ a displaced vertex trigger, LHCb \cite{ReOpt} is designed to
trigger on and reconstruct such events.
LHCb operates at a center of mass energy equal to that
of ATLAS and CMS (14 TeV).  However, its luminosity is limited to 2 fb$^{-1}$
per year and it covers only the forward region.  The experiment is designed to make measurements of
rare $b$-hadron decays by relying on their ability to precisely
reconstruct displaced vertices.  The lower luminosity limits the average number of interactions
per bunch crossing to $\lsim$ 1, which allows for more precise
vertexing.  This makes it an ideal experiment to search for our signal.

The purpose of this article is to show quantitatively that the LHCb
experiment should have significant reach in the parameter
space of this class of supersymmetric models.  In parts of parameter
space, it may be able to show the first direct evidence of the
lightest neutralino and the lightest Higgs boson.  Below we present our
estimates of the LHCb's physics reach with regard
to squark and Higgs production.  Our work shows that the signal events
easily pass the lowest level LHCb triggers.  We suggest a modified
high level trigger to increase the efficiency with which 
the signal is written to tape.
While computational limits prohibit our complete understanding of the
leading order QCD background, we argue that for some parts of parameter
space the signal will dominate the background.

\section{Neutralino Decay via Baryon Number Violation}
The baryon number violating operators in Eq. (\ref{eq:BNV}) involve
nine complex couplings (because $j$ and $k$ are anti-symmetric).  When
the neutralino decays, it does so via the coupling $\lambda''_{ijk}$
into up-type quark $i$ and down-typed quarks $j$ and $k$ through an
off-shell squark.  A reasonable, theoretically motivated
parameterization for these couplings based on a spurion analysis of
flavor breaking in the standard model is \cite{Hinchliffe:1992ad}
\begin{equation}
\lambda''_{ijk} = \lambda''_0 \sqrt{\frac{m_i m_j m_k}{v^3 \sin\beta \cos^2\beta}},
\end{equation}
where the $m_i$, etc., are quark masses 
\cite{note1}, and $v \sin\beta$ and $v \cos\beta$ are the vacuum
expectation values of the up-type and down-type Higgses
respectively, with $v=174$ GeV.  We shall use this parameterization and take
$\tan\beta=1$ since any difference can be absorbed into $\lambda_0$.
Note, the $\lambda''_{323}$ coupling dominates.  The dominant decay
mode of the neutralino will be $\chi\rightarrow tbs$, unless the
neutralino is lighter than the top in which case $\chi\rightarrow
cbs$ dominates.  
%The idea of this parameterization is that standard model Yukawa couplings (and thus quark masses) break the underlying 
%flavor symmetry of the gauge interactions, and thus the new couplings should break the same symmetry by in some sense 
%the same amount (with the same spurions).  The overall coefficient $\lambda''_0$ is a potential additional suppression 
%(or enhancement) with parameterizes our ignorance of how R parity is violated.  
In either case, neutralino decays are dominated by heavy flavors, and should contain additional displaced
vertices.  We will not utilize this additional handle on the signal,
though it may prove to be a useful part of the full experimental analysis.

The strongest current bounds on the magnitude of $\lambda''$ couplings
are from baryon number violating processes, namely neutron-antineutron
oscillations and double nucleon decay in, for example, oxygen nuclei
\cite{Dimopoulos:1987rk}.  Such bounds allow
$\lambda''_0 \sim {\cal O}(1)$ within QCD uncertainties.  If the
$\lambda''$ have arbitrary complex phases, they can contribute to
direct CP violation in Kaon decays and to $K - \bar{K}$ mixing.  The
strongest bound in this case is the limit ${\cal I}(\lambda''_{313}
\lambda''_{323}*) < 10^{-8}$ \cite{Barbieri:1985ty}, which implies a
bound on our universal parameter $\lambda''_0 \lsim 1/20$ if all phase
differences are order unity and squark masses are 100 GeV.  There are
no significant bounds on the individual $\lambda''_{223}$ and
$\lambda''_{323}$ couplings.  For a complete review, see \cite{Barbier:2004ez}.

The proper lifetime of the neutralino depends on the R-parity
violating couplings, the neutralino mass $m_\chi$, and the squark
masses $m_{\tilde q}$.  With the simplifying assumptions of a
universal squark mass at low energies and a single dominant R-parity
violating coupling (as in our parameterization), the proper lifetime is
\begin{eqnarray}
\tau_\chi & \simeq & \frac{384\pi^2 \cos^2\theta_w}{\alpha \left| U_{21}\right|^2 \lambda''^2}\frac{m_{\tilde{q}}^4}{m_\chi^5}% ( \beta\gamma)
\\\nonumber
& \sim & \frac{3 {\mu}{\rm m}}{c \left| U_{21}\right|^2} \left(\frac{10^{-2}}{\lambda''}\right)^2 \left(\frac{m_{\tilde q}}{100\; {\rm GeV}}\right)^4 \left(\frac{30\; {\rm GeV}}{m_\chi}\right)^5 .%\frac{p_\chi}{m_\chi},
\end{eqnarray}
where $|U_{21}|$ is an element of the neutralino rotation matrix (see
\cite{Carpenter:2006hs}).  We have neglected Yukawa couplings, QCD
corrections and phase-space corrections (taking final state particles
as massless).  These are good approximations in the two cases we
study.  Yukawa couplings are relevant to the extent that the lightest
neutralino is partially higgsino.  For Higgs production and decay, the
neutralino is much lighter than the top, and for the decay to
neutralinos to dominate, it turns out $\tan\beta$ should not be too large
\cite{Carpenter:2006hs}, and therefore all relevant Yukawa couplings
are small.  In the case of squark production, we will look only at the
`pure bino' limit (making the higgsinos and winos heavy), and thus we
can ignore the Yukawas entirely.  In the former case, $|U_{21}|$ is
less than and of order unity.  In the latter case (pure bino limit), $|U_{21}|=1$.

\begin{figure}
\resizebox{\hsize}{!}{\includegraphics{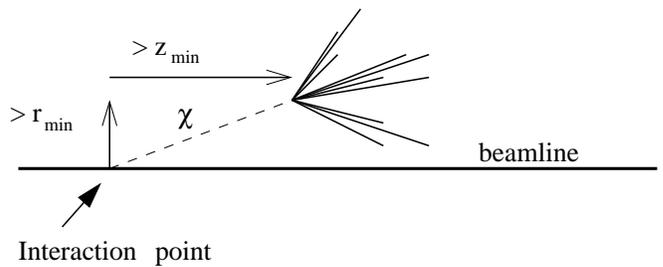}}
\caption{A qualitative picture of a neutralino decay off the beamline.
The decay is of a neutral particle (no track) into a large multiplicity 
of tracks.  The invariant mass of the tracks should be significantly larger
than those from a $b$-hadron decay.  The quantities $r_{min}$ and $z_{min}$ are
defined in the text.}
\label{fig:decay}
\end{figure}

\section{SIGNALS AND BACKGROUNDS AT LHCb}
Here we list the expected signals and backgrounds, proposed triggers
and signal efficiencies, and offline discriminants:
\begin{itemize}
\item In Figure \ref{fig:decay} we show pictorially a macroscopic decay of 
a neutralino off the beam line.  The production signals we study are:
  \begin{itemize}
  \item Squark production with $\tilde{q} \rightarrow q \chi_0$ and $\chi_0 \rightarrow qqq$. 
  \item Higgs production with $ h \rightarrow \chi_0 \chi_0$ and $\chi_0\rightarrow qqq$.  
  \end{itemize}
  They are generated, including showering, with Pythia
  v6.400 \cite{Pythia}.  For squark production we use the following parameters:
  \begin{itemize}
  \item The ratio of couplings 
    $(\lambda^{''}_{223}/\lambda^{''}_{323})=(1/20)$.
  \item A scan of $m_{\tilde{q}}$ from 100-1000 GeV in 100 GeV steps.
  \item A scan of three different bino masses: $M_{1} =$ 40, 100, 200 GeV and 
	three different coupling values ${\lambda^{''}_{223}}=$ $10^{-3}$,
        $10^{-4}$,  and $10^{-5}$.  
  \item $M_2 = M_3 = \mu = 1.2$ TeV while all other parameters 
        are set to the Pythia default values.  
  \end{itemize}
For Higgs
  production we do the same scan of $M_1$ parameters and use $\tan\beta=5$, 
  $M_2=250$ GeV, $\mu=120$ GeV,
  $m_{\tilde{q}}= A_t = 1$ TeV and $\lambda^{''}_{223}=10^{-2}$, with other soft terms at default values
  and other $\lambda''$ couplings set to zero. 
\item The background is taken to be multiple $b$ production.  We use
  Pythia to simulate $b\overline{b}$ events.  Madgraph v4.1.19 \cite{Madgraph} is used
  to compute matrix elements of $gb \rightarrow$ $bb\overline{b}$,
  $gg\rightarrow$ $b\overline{b}b\overline{b}$, and $gg\rightarrow$
  $b\overline{b}c\overline{c}$ while Pythia is
  used to shower these events.  
\item We find that the following cuts and triggers discriminate the signal
  and background:
  \begin{itemize}
  \item Requiring a displaced vertex with 300 $\mu$m $< z <$ 0.4 m and $ r > $ 60 $\mu$m, where $z$ and $r$ are the horizontal and perpendicular distance from the interaction point. 
  \item Requiring at least 5 tracks from the displaced vertex.
  \item Requiring at least two tracks with $p_T \geq$ 1 GeV and a two-dimensional impact
    parameter of 0.07 mm $< b_{IP} <$ 15.0 mm
  \end{itemize}
\item For offline discrimination we use invariant mass distributions
  of displaced vertices.

\end{itemize}
We now describe the relevant aspects of the LHCb experiment \cite{ReOpt}
and explain in
detail the motivation for and expected results of these cuts.
LHCb is asymmetric in theta (polar angle) acceptance.  The
horizontal acceptance is 15 mrad $< \theta <$ 300 mrad while the
maximum vertical acceptance is 250 mrad.  For simplicity we restrict
our analysis to the region 15 mrad $< \theta <$ 250 mrad.  Offline
reconstruction of the primary vertex is expected to have a resolution
of $\lsim$ 50 $\mu$m along the beam line and $\lsim$ 10 $\mu$m
perpendicular to it .  The typical $z$ resolution of a secondary vertex is
$\sim$ 200 $\mu$m.  Transverse resolution is $p_{T}$ dependent, and is
$\sim$ 20 $\mu$m for 1 GeV $p_{T}$ track.  We assume that vertexing may
be done up to 0.4 m along the beamline which corresponds to half of the
Vertex Locator length \cite{schneider},
%(though in principle it may be possible to reconstruct vertices displaced up to a few meters,)
We set the resolution of a displaced vertex to be a cylinder of 200 $\mu$m in $z$ and
30 $\mu$m in $r$.  This means that if a second vertex lies outside this cylinder then it can be distinguished,
otherwise it cannot.  We denote these lengths as $\delta z$ and
$\delta r$.  The required minimum distances from the primary vertex as
described above in $z$ and $r$ are denoted $z_{min} =$ 300 $\mu$m and
$r_{min} =$ 60 $\mu$m.  No 
detector effects beyond vertex resolution are considered.

\subsection{Squark Production}
All superpartners produced at the LHC cascade to the lightest
neutralino (direct decays via R-parity violation are suppressed by the
small $\lambda''$ coupling).  For simplicity we look at squark pair
production where each squark decays to a quark and the lightest
neutralino.  The goal of this search is to see one of these
neutralinos in the LHCb acceptance.  The signal we look for is a
displaced vertex with a larger track multiplicity than a typical $b$
decay and an invariant mass of the tracks larger than the $b$ mass.
The decay length may or may not be similar to a typical $b$ hadron, so
we do not use this as a distinguishing feature. 
%% \par
%% We use Pythia v6.0x \cite{Pythia} to simulate the both the squark and
%% Higgs signal.
%% In the squark case we set the couplings the ratio of
%% $\frac{\lambda^{''}_{223}}{\lambda^{''}_{323}}=\frac{1}{20}$ and
%% scan $m_{\tilde{q}}$ from 100-1000 GeV for the points
%% $m_{\chi_{0}} =$ 38, 98, 198 GeV and ${\lambda^{''}_{223}}=$ $10^{-3}$,
%% $10^{-4}$, $10^{-5}$. $M_2$ and $M_3$ are set to 1.2 TeV. 
\begin{figure}
\resizebox{\hsize}{!}{\includegraphics{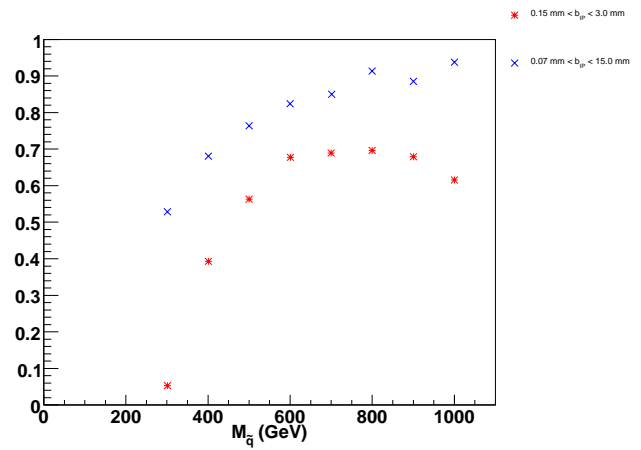}}
\caption{Efficiency of $\chi_0$ acceptance with respect to impact
parameter requirements.  We require 5 tracks, 2 of
  which with $p_T$ $>$ 1 GeV.  The lower points show 0.15 mm $<$
  $b_{IP}$ $<$ 3.0 mm and the upper points show 0.07 mm $<$ $b_{IP}$
  $<$ 15.0 mm.  The parameters are set at $\lambda^{''}_{223}=10^{-4}$, 
  $M_1 =$ 200 GeV, and $m_{\tilde q} = 400$ GeV.} 
\label{fig:impactplot}
\end{figure}

LHCb's Level 0 (L0) trigger is designed to reject multiple primary vertices
(`pile up' events) and events with large numbers of tracks (`busy'
events). % We postpone the discussion of High Level Triggers until
% after examining the offline signal features.
L0 reduces the data rate from $\sim$ 12 MHz to $\sim$ 1 MHz. 
We find that 95$\%$ of our squark production signal (one neutralino
leaving at least 5 tracks in the detector acceptance) passes L0. 
The High Level Triggers (HLT) are responsible for
reducing the rate to 2 kHz, the read out rate.  One component of the HLT is 2D track
reconstruction searching for tracks with high $p_T$ ($\gsim$ 1 GeV) and large
impact parameter, 0.15 mm $<$ $b_{IP}$
$<$ 3.0 mm, tracks.  Our signal generically produces more high $p_T$ tracks than the
background because of the neutralino's greater mass and because it is the product of
a heavy particle decay.  However, we find that the proposed impact parameter 
window results in signal efficiencies below 10$\%$ for decay lengths
inconsistent with that of a $b$.  The signal efficiency is increased to
above 50$\%$ in most parts of parameter space if the impact parameter window is widened to 0.07 mm $<$
$b_{IP}$ $<$ 15.0 mm.  Figure ~\ref{fig:impactplot} shows the
efficiencies of the two ranges for a particular point in parameter space.

Extending the impact parameter range to a lower value of 50 $\mu$m is suggested in the
context of LHCb upgrades \cite{Muheim:Beauty06}.  The feasibility of
extending the range to large values in unknown and requires a detector
simulation.  The naive background for large impact parameter tracks is
strange decays.  This is because $\tau_{strange} c \sim$ centimeters.  If this
is the case, it seems plausible that requiring multiple tracks with
high $p_{T}$ can significantly reduce this background.  Henceforth we
optimistically assume that the range can be extended.  

\begin{figure}
\resizebox{\hsize}{!}{\includegraphics{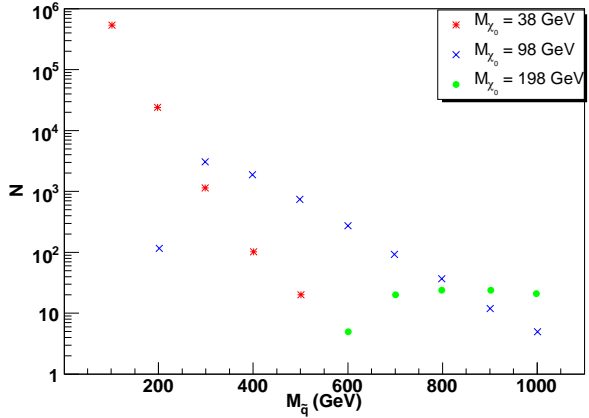}}
\caption{Number of expected $\chi_0$ events from squark production
  vs. squark mass.  At least
  5 tracks with 2 having more than 1 GeV of $p_{T}$ and 0.07 mm $<$
  $b_{IP}$ $<$ 15.0 mm are required.  The coupling is $\lambda^{''}_{223}=10^{-4}$ and the neutralino masses are computed by Pythia using the parameters set at the beginning of the section.}
\label{fig:acceptcp1}
\end{figure}

\begin{figure}
\resizebox{\hsize}{!}{\includegraphics{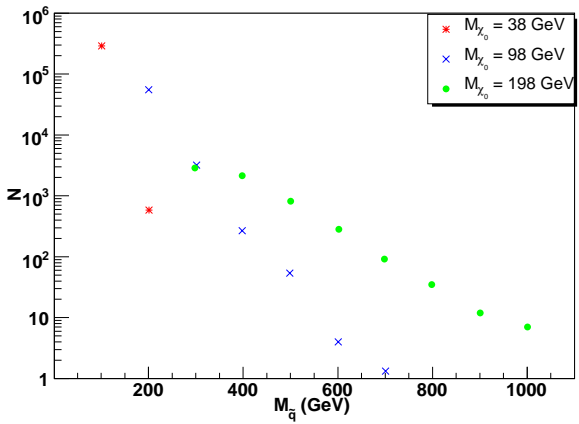}}
\caption{The same as Fig ~\ref{fig:acceptcp1} but with $\lambda^{''}_{223}=10^{-5}$} 
\label{fig:acceptcp2}
\end{figure}

The expected
event rate of neutralinos from squark production that pass our cuts
are shown in Figures ~\ref{fig:acceptcp1} and ~\ref{fig:acceptcp2}. 

\begin{figure}
%\centering
\resizebox{\hsize}{!}{\includegraphics{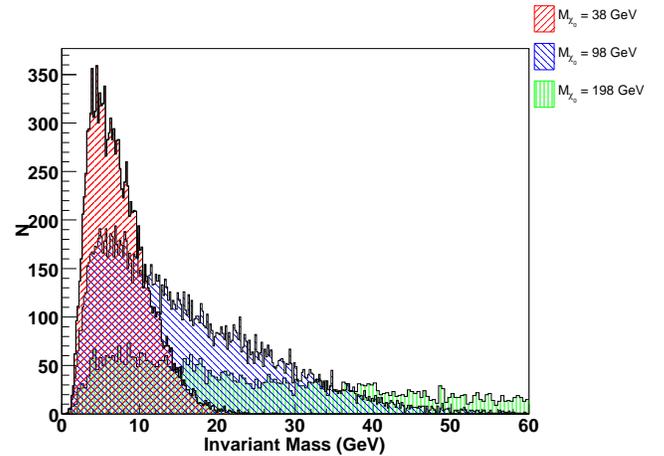}}
\caption{$\chi_0$ Invariant Mass from squark production.  All points are
  $\lambda^{''}_{223}=10^{-5}$ with the same requirements as Figure
  ~\ref{fig:acceptcp1}.  Red: $M_{1}=40$
  GeV, $m_{\tilde{q}}=100$ GeV.  Blue: $M_{1}=100$
  GeV, $m_{\tilde{q}}=200$ GeV.  Green: $M_{1}=200$ GeV, $m_{\tilde{q}}=400$ GeV.} 
\label{fig:squarkmass}
\end{figure}

%% The expected signal is large invariant of tracks from a displaced vertex.
Figure \ref{fig:squarkmass} shows the signal invariant mass distributions 
of tracks for events that pass our cuts. 
As expected, these distributions reach far beyond the mass of a 
$b$-hadron, and thus we need not worry about isolated $b$ decays as a background.  We find rather that the most important 
background (offline) consists of
$b$-hadron events in which another particle, produced at the primary
vertex, decays near enough to the $b$ that the vertices cannot be
resolved individually by the detector.  We refer to these as
overlapping events.  

\begin{figure}
%\centering
\resizebox{\hsize}{!}{\includegraphics{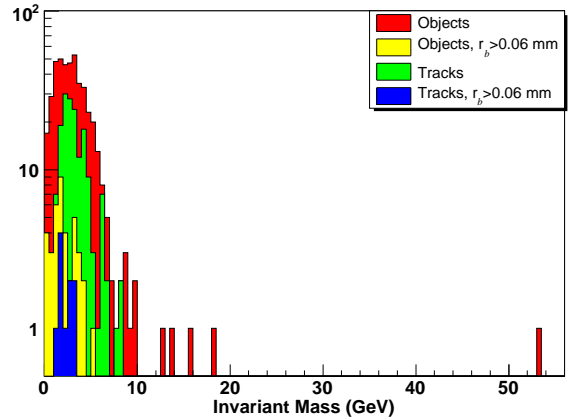}}
\caption{Background invariant mass of overlapping events in 2$b$ and 3$b$ 
  production.  Objects are defined as all
  decay products that deposit energy in the detector (\textit{i.e.} everything but
  $\nu$'s).  We demand more than 5 objects or more than 5 tracks from a vertex.  No
  cuts are made on impact parameter nor track$/$object $p_T$.  All
  large mass events in red are due to $b$ decay overlapping with a non$-b$ decay.} 
\label{fig:background}
\end{figure}

We now give a rough quantitative estimate of the background.  Our region of
interest is for track invariant masses above $2 m_B \sim 12$ GeV. We see
from Figure \ref{fig:squarkmass} this region has a significant overlap with our signal.  Our limited computing power only allows us to simulate $10^{-5}$
years of background, in which we find no events which pass our track cuts and have an invariant
mass above 12 GeV (see Figure \ref{fig:background}).  To better
understand the background, we also look at the invariant mass of all
decay products (charged and uncharged) from these overlapping events.
Using this information, we are able to define cuts which should in
principle reduce our background to less than 1000
 events per year.  Below, we describe how we come to this estimate.
\par 
The expression for the invariant mass of two particles is
\begin{equation}
M^2=m^2_{1}+m^2_{2}+2(E_{1}E_{2}-p_{z_{1}}p_{z_{2}}-p_{T_{1}}p_{T_{2}}cos\Delta\phi).
\label{eq:Mass}
\end{equation}
where $m_i$, $E_i$, $p_{z_i}$, and $p_{T_i}$ are the mass, energy, $z$-momentum and transverse
momentum respectively of the $i$th particle, and $\Delta\phi =
\phi_1-\phi_2$ is the difference in the azimuthal angle of the two
particle momentum vectors.  There are two overlap cases: 2$b$s and a $b$ plus a non-$b$.  We discuss the 2$b$ case.  It is clear
that the non-relativistic
limit cannot produce $M^2 \gg (2m_b)^2$.  The relativistic limit reduces
(\ref{eq:Mass}) to 
\begin{eqnarray}
M^2  & \simeq &2m_b^2+2p_{z_1}p_{z_2}\left(\frac{1 - c_1 c_2 - s_1 s_2 \cos\Delta\phi}{c_1 c_2}\right)\nonumber\\ & + & m_b^2
\left(\frac{p_{z_1} c_2}{p_{z_2} c_1} +\frac{p_{z_2} c_1}{p_{z_1} c_2}\right)\label{eq:Mass2}
\end{eqnarray}
where $c_i \equiv \cos\theta_i$, the cosine of the polar angle of the
momentum vector of the $i$th particle, and similarly,
$s_i\equiv\sin\theta_i$.  (Note, the region of parameter space where
one $b$ is non-relativistic is a special case of what we discuss
below). 

Examining the cross term we see that there are
two interesting cases: $p_{z_1} \sim p_{z_2}$ and $p_{z_i} \gg
p_{z_j}$.  The former case requires a large difference in polar or azimuthal angles
to generate a large cross term.  Maximizing this difference (for example, in the polar 
angle) while demanding a large cross term and using $p_T=p_z\tan\theta$ leads to a
minimum $p_T$ for the $b-$hadrons.  Furthermore a large $\theta$ difference with a
small transverse distance (making the vertices unresolved)
requires the vertices to be as near the primary
vertex as possible.  These considerations significantly suppress the number of
overlapping events.  As an illustration we take $\delta r =30 \mu$m, the closer
$b$ a transverse distance of 30 $\mu$m from the $z$ axis, and the vertices a
distance of 360 $\mu$m from the primary vertex along the $z$ direction.  Demanding 
the cross term give 16 $m_b^2$ (to get an invariant mass of all decay products, not just 
tracks, of just over 20 GeV) we
find that $p_z \gsim$ 160 GeV.  This corresponds to $p_T \gsim$ 20 GeV
for the softer $b$, a requirement which suppresses the cross section by better than
$10^{-5}$ and makes this parameter range irrelevant.
\par
Conversely, the case in which $p_{z_i} \gg p_{z_j}$ is important even when the
$\theta$ difference is small.  The non-relativistic corrections
-- the last term in Eq. (\ref{eq:Mass2}) -- dominate when the
difference in angles vanish.  Generating a cross term of $ 16 m_b^2$
requires a ratio of 16:1 between the $p_z$'s.  The softer $b$ (call it
$b_1$) decays dominantly at a length $L\lsim(p_{z_1}/m_b)  
\tau_b c$.  Now the harder $b$ has $p_{T_2} \simeq p_{z_2}\theta \simeq
16 p_{z_1} \theta$.  Using the requirement that $\theta \geq
\frac{r_{min}}{L_1}$ (so the displaced vertex satisfies our $r_{min}$ cut), and plugging the values of $r_{min}$ and $ \tau_b c$ leads to
the requirement $p_{T_2}\gsim 2m_b$.  To estimate our background, we create
a sample of $10^{-5}$ years of 2$b$ production using Pythia (roughly
$10^7$ events at leading order) requiring both $b$s to decay within
the acceptance of our detector and to pass our $r_{min}$ and 
$z_{min}$ requirements.  We then count the number of events that satisfy
$(p_{z_2}/p_{z_1}) \ge 16$ and $p_{T_2}\geq 10$ GeV.  The fraction
of our sample which passes these cuts is one part in  $2\times
10^{3}$.  Then we take the same sample without the momentum
requirements and find the number of overlap events to be 69 -- or
scaled up, roughly $7\times 10^6$ per year.  If we take the
distribution of momenta among these events to be flat (overly
conservative), we can simply take a product of the two suppressions
% the two different requirements pose on our sample 
and estimate the number
of events which have the potential to have a large enough invariant
mass.  Our estimate is $N < 7\times 10^6 \times 5 \times 10^{-4} =
3,500$.  If we include the fact that $b$'s with very different momenta
will have very different decay lengths, we find another suppression of
a factor of nearly an order of magnitude and thus expect a background
to be at most on the order of hundreds of events. 
%% we see that there are important two limit of Eq. \ref{eq:Mass}:
%% cos$\Delta \phi$ $<$ 0, $p_{z_i} >> m_i$ and
%% $p_{z_i} >> m_i$, $m_j \gsim p_{z_j}$. In the former case the $p_{T}$
%% term dominates and adds to $M^2$. This is manageable; qualitatively
%% because LHCb is a low $p_T$ machine and quantitatively because such kinematics
%% can be eliminated by demanding the cos$\Delta \phi$ $>$ 0. The
%% requirement is that vertices must have
%% displacement in R greater than the detector resolution, $\delta r$. The detector
%% geometry is such that this constrain only affects z $<$ tan$\theta_{min}
%% * \delta r$ $\sim$ 90*30 $\mu$m $=$ 2.7 cm. Although this is a small 
%% portion of the vertexing region a significant fraction of
%% overlapping events occur here because the z resolution,$\delta z$ $=$
%% 200 $\mu$m, is comparable
%% to this scale. The kinematic case $p_{z_i} >> m_i, m_j \gsim p_{z_j}$
%% presents no obvious cut that could remove it. 
%% %% One might consider imposing a cut on the variance of track $p_z$ to
%% %% eliminate such event, however it is unfruitful because the signal
%% %% generically produces particles with a large spread in $p_z$. 
%% However, we find no overlapping events with such
%% kinematics. Furthermore, we find no overlapping events that contain
%% more than two particles. This is significant because track and
%% invariant cuts clearly become less distinguished between the
%% background and the signal when many particles overlap.

In addition to the $b\bar{b}$ background, there are overlap events generated
in, for example, the 3$b$ background.  We find no events where two $b$s overlap
giving a large invariant mass, and using similar arguments to those above find that 
a full year should produce at most as many events as in the 2$b$ sample.  However, we
do find large invariant mass events in this sample which involve the overlap of a $b$
and a strange hadron decay.  These events do not pass either 
the 5 track cut or the $r_{min}$ requirement.  Imposing both should in principle severely 
limit or eliminate events of this type, but unfortunately it is difficult to estimate.  We
will simply assume they can be removed by these or similar cuts.  In
Figure \ref{fig:background} we plot the invariant mass of overlapping vertices in the 
3$b$ sample.  We include in the plot the invariant mass of all decay products to see the
large invariant mass events.
All of the events with invariant masses larger than 10 GeV are due to
$b$-non-$b$ overlapping events.

The other
simulated backgrounds produce
an overlapping event fraction about that of $b\bar{b}$, and they are
cross section suppressed by more than an order of magnitude.

\par
\par
We now estimate the parameter space that can be explored by LHCb.  We
assume a background of 400 events above 12 GeV of track invariant mass.
Significance at the level of $\frac{S}{\sqrt{B}} > 5$
requires $\gsim$ 100 signal events above 12 GeV.  The regions of parameter 
space which exceed this event rate after cuts are:
%%  Referring to Figure
%% ~\ref{fig:squarkmass} and Figures ~\ref{fig:acceptcp3}
%% ~\ref{fig:acceptcp1} ~\ref{fig:acceptcp2} this leaves the following:
\begin{itemize}
\item $\lambda^{''}_{223}=10^{-3}$: 
  \begin{itemize}
  \item $M_{\chi_0}= 38$: 200 GeV $\lsim$ $m_{\tilde{q}}$  $\lsim$ 600 GeV. 
  \end{itemize}
\item $\lambda^{''}_{223}=10^{-4}$:
  \begin{itemize}
  \item $M_{\chi_0}= 38$: 100 GeV $\lsim$ $m_{\tilde{q}}$  $\lsim$ 400 GeV. 
  \item $M_{\chi_0}= 98$: 200 GeV $\lsim$ $m_{\tilde{q}}$  $\lsim$ 700 GeV. 
  \end{itemize}
\item $\lambda^{''}_{223}=10^{-5}$:
  \begin{itemize}
  \item $M_{\chi_0}= 38$: 100 GeV $\lsim$ $m_{\tilde{q}}$  $\lsim$ 200 GeV. 
  \item $M_{\chi_0}= 98$: 200 GeV $\lsim$ $m_{\tilde{q}}$  $\lsim$ 400 GeV. 
  \item $M_{\chi_0}= 198$: 300 GeV $\lsim$ $m_{\tilde{q}}$  $\lsim$ 700 GeV. 
  \end{itemize}
\end{itemize}
%%%%%%%%%%%%%%%%%%%%%%%%%%%%%%%%%%%%%%%%%%%%%%%%%%%%%%%%%%%%%%%%%%%%%
%%NOTES:
%% m_chi= 38 => 3133/13480 ~ 25% of invariant mass distribution above
%% 10 GeV, 1705/13480 ~ 13% above 12 GeV, 553/13840 ~ 4% above 15 GeV

%% m_chi = 98 => 11620/17525 ~ 65% above 10 GeV, 9743/17525 ~ 55%
%% above 12 GeV, 7868/17525 ~ 45% above 15 GeV

%% m_chi = 198 => 3061/5246 ~ 60% above 10 GeV, 2675/5246 ~ 50%
%% above 12 GeV, 2193/5246 ~ 40% above 15 GeV
%% these %'s are smaller than 98 GeV b/c binning is cutoff at 60 GeV
%%%%%%%%%%%%%%%%%%%%%%%%%%%%%%%%%%%%%%%%%%%%%%%%%%%%%%%%%%%%%%%%%%%%%

\subsection{Higgs Production}
\par
There exists an interesting region of parameter space in
which the Higgs dominantly decays to neutralinos.  The signal invariant mass
distribution becomes more background-like in this parameter space as
$m_{\chi_{0}}$ becomes lighter. 
In fact the distribution is almost
indistinguishable from the overlap background in Figure \ref{fig:background}
for $m_{\chi_0} \lsim$ 20 GeV, in part because the neutralino vertex loses
some of its track invariant mass through its decay to a $b$ whose decay 
products often reconstruct at a different point.
\begin{figure} 
%\centering
\resizebox{\hsize}{!}{\includegraphics{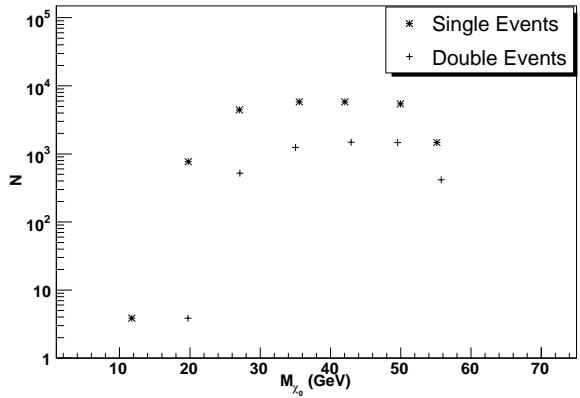}}
\caption{Number of Expected $\chi_0$ in the acceptance from Higgs production
  and decay at
  2 fb$^{-1}$.  We require 5 tracks, 2 of which with more than 1 GeV of
  $p_T$ and 0.07 $<$ $b_{IP}$ $<$ 15.0 mm.  This point is $\tan\beta = 5$,
  $M_2 =250$ GeV, $\mu =120$ GeV, $m_{\tilde q} = A_t =1$ TeV,
  and all other parameters at Pythia default values.  The mass of the Higgs at this point is $m_h\sim 115$ GeV.} 
\label{fig:higgs_accept}
\end{figure}

However, a distinguishing characteristic of the signal is the possibility
of both neutralinos being in the acceptance.  An exemplary point is
shown in Figure \ref{fig:higgs_accept} where we use the leading order
inclusive Higgs production cross section of $\sigma_{h}=
20$ pb \cite{kirill}.  We see that at this point if $m_{\chi_{0}} \gsim 25$ GeV
then there are a significant number of Higgs decays that deposit
both neutralinos in the detector.  The expected background for this
signal is two sets of overlapping decays which should be negligiable.

\par
The rapid fall off of the
acceptance distribution at small $m_{\chi_0}$ 
%% is a combination of the light neutralino mass relative to $m_h$ and 
is primarily due to long decay length at these
particular points. 
%% In the Higgs frame the neutralinos are back to back
%% and thus lighter neutralinos are more boosted in this frame as compared
%% to heavier neutralinos. It follows that in order for two light
%% neutralinos to be in the LHCb acceptance requires a more boosted Higgs in
%% the lab frame as compared to the case of two heavy neutralinos. For
%% the masses of interest $m_h \sim$ 120 GeV and $m_{\chi_0} \gsim$ 10
%% GeV \footnote[4]{The neutralino mass can't be smaller if it is to
%%   decay to bcs} the neutralino has a boost factor of $\sim$ 6 and thus
%% the Higgs must have $E_h \gsim$ 700 GeV to counteract the decay
%% kinematics. WANT AN ARGUMENT THAT SAYS PRODUCING A 700 GEV HIGGS IN
%% FOWARD DIRECTION ISN'T OBSCENELY DIFFICULT, IE THE DROP IS DECAY
%% LENGTH LIMITED. 
The decay length of a 20 GeV neutralino for the
prescribed parameters is $L \sim$ 0.2 m.  Given that such a neutralino is
boosted by at least a factor 3 due to the Higgs decay, we would not
expect both neutralinos to be in the acceptance (defined with an upper
z limit of 0.4 m) whereas it is not
surprising that one of the decays is in the detector.  Reducing
$m_{\chi_0}$ by a factor of 2 leads to 32 fold increase, plus 
a significant phase-space increase of $L$ -- hence
the sparseness of accepted events at small $m_{\chi_0}$.

\section{DISCUSSION}
%The most studied scenario of New Physics discovery at the LHC is
%mSUGRA at either ATLAS or CMS. The theoretical motivation for mSUGRA
%is momentous, but empirical evidence forces consideration of less
%constrained supersymmetric models. One such model is the simple
%addition of Baryon Number Violating operator to the MSSM
%superpotential. The phenomenology of this model 
The addition of baryon number violating operators to the MSSM
superpotential allows for a more natural model of supersymmetry while
also producing phenomenology that may pose difficulties for
ATLAS and CMS.  The central phenomenological signature is
displaced vertices for which the LHCb is well suited to observe and
reconstruct.  Displaced vertices of $b-$hadron decays is a potentially
enormous background.  Despite computational limitation that forbid a
full simulation of the background we can argue that it is plausible for
large portions of parameter space to be explored.  This is a
consequence of the large invariant mass distribution of our signal,
that necessitates coincident background decays.  Thus we estimate
that LHCb could rule out a significant portion of parameter space.  
However, only the most na\"{i}ve detector issues have been considered and a full
detector simulation is needed to understand the detector's true reach.

In addition to R-parity violating supersymmetry, other versions of supersymmetry
may also contain displaced vertices.  This includes parts of parameter space with 
near degeneracies between the LSP and NLSP which could occur between, for example, 
a stau and a neutralino, or between neutralinos in theories
with an added singlet field.  Finally, so-called `hidden valley' models 
\cite{Strassler:2006im} 
also give rise to non-standard displaced vertices and have been suggested as good candidates for
LHCb physics.  A dedicated search at LHCb may provide the first discovery of new physics at the LHC. 
%We conclude a dedicated search for new physics at LHCb is certainly warranted.

\vskip 0.8cm

We thank Aurelio Bay, Olivier Schneider, and Frederic Teubert for useful discussions
and feedback, and especially  Petar Maksimovic for pointing us towards LHCb. 
This work is supported in part by NSF grants PHY-0244990 and PHY-0401513, 
by DOE grant DE-FG02-03ER4127, and by the Alfred P. Sloan Foundation.


\begin{thebibliography}{1}
\bibitem{note1}
In principle, a top-down theoretical model for these
  couplings would associate their values to quark masses at high
  scales and then run the couplings down to the neutralino mass.
  Practically speaking, one should using the running quark masses at
  the neutralino mass (or roughly the weak scale) which results in
  order unity differences from their current masses.
  Pythia \cite{Pythia}, which does implement this parameterization for
  R-parity violating couplings as an option, unfortunately uses the
  constituent quark masses for the light quarks, which has nothing to
  do with the flavor breaking. This option in Pythia should not be used.
\end{thebibliography}
\end{document}